\documentclass{sig-alternate-10pt}
\usepackage[margin=1in]{geometry}
\usepackage{graphicx}
\usepackage{url}
\usepackage[flushleft]{threeparttable}

\begin{document}
	\pagestyle{empty}
	
	\title{Database Meets Deep Learning: Challenges and Opportunities}
	\numberofauthors{1}
	\author{
		\alignauthor
		Wei Wang$^\dag$,  Meihui Zhang$^\ddag$, Gang Chen$^\S$, \\ H. V. Jagadish$^\#$, Beng Chin Ooi$^\dag$, Kian-Lee Tan$^\dag$\\
		\affaddr{$^\dag$National University of Singapore \quad $^\ddag$ Beijing Institute of Technology\\
			$^\S$Zhejiang University \quad $^\#$University of Michigan}
		\email{$^\dag$\{wangwei, ooibc, tankl\}@comp.nus.edu.sg \quad$^\ddag$meihui\_zhang@bit.edu.cn $^\S$cg@zju.edu.cn \quad$^\#$jag@umich.edu}
	}
	
	\maketitle
	
\begin{abstract}

	Deep learning has recently become very popular on account of its incredible success in many complex data-driven applications, such as image classification and speech recognition.
	The database community has worked on data-driven applications for many years, and therefore should be playing a lead role in supporting this new wave. However,
	databases and deep learning are different in terms of both techniques and applications.
	In this paper, we discuss research problems at the intersection of the two fields.
	In particular, we discuss possible improvements for deep learning systems from a database perspective, and analyze database applications that may benefit from deep learning techniques.
\end{abstract}
	
	\section{Introduction}
	In recent years, we have witnessed the success of numerous data-driven machine-learning-based applications. This has prompted the database community to investigate the opportunities for integrating machine learning techniques in the design of database systems and applications~\cite{DBLP:conf/sigmod/ReABCJKR15}.
	A branch of machine learning, called deep learning~\cite{lecun2015deep,Goodfellow-et-al-2016-Book},
	has attracted worldwide interest in recent years due to its excellent
	performance in multiple areas
	including speech recognition,
	image classification
	and natural language processing (NLP).
	The foundation of deep learning was established about twenty years ago in the form of neural networks.
	Its recent resurgence is mainly fueled by three factors: immense computing power,  which reduces the time to train
	and deploy new models, e.g. Graphic Processing Unit (GPU) enables the training systems to run much faster than those in the 1990s;
	massive (labeled) training datasets (e.g. ImageNet) enable a more comprehensive knowledge of the domain to be acquired; new deep learning models (e.g. AlexNet~\cite{krizhevsky2012imagenet}) improve the ability to capture data regularities.

	Database researchers have been working on system optimization and large scale data-driven applications since 1970s, which are closely related to the first two factors. It is natural to think about the relationships between databases and deep learning.
	First, are there any insights that the database community can offer to deep learning?
	It has been shown that larger training datasets and a
	deeper model structure improve the accuracy of deep learning models.
	However, the side effect is that the training becomes more costly. Approaches have been proposed to accelerate the training speed from both the system perspective~\cite{DBLP:conf/icml/CoatesHWWCN13, jia2014caffe, Dean2012, opensourcesinga, tensorflow2015-whitepaper} and the theory perspective~\cite{zeiler2012adadelta,gao2015active}.
	Since the database community has rich experience with system optimization, it would be opportune to
	discuss the applicability of database techniques for optimizing deep learning systems.
	For example, distributed computing and memory management are key database technologies also central to deep learning.

	Second, are there any deep learning techniques that can be adapted for database problems? Deep learning emerged from the machine learning and computer vision communities.
	It has been successfully applied to other domains, like NLP~\cite{DBLP:journals/corr/Goldberg15c}.
	However, few studies have been conducted using deep learning techniques for traditional database problems.
	This is partially because traditional database problems --- like indexing, transaction and storage management --- involve less uncertainty, whereas deep learning is good at predicting over uncertain events.
	Nevertheless, there are problems in databases like knowledge fusion~\cite{DBLP:journals/pvldb/DongGHHMSZ14} and crowdsourcing \cite{DBLP:journals/sigkdd/OoiTTY0LNTZ14}, which are probabilistic problems. It is possible to apply deep learning techniques in these areas.
	We will discuss specific problems like querying interface, knowledge fusion, etc. in this paper.
	
	The previous version~\cite{Wang:2016:DMD:3003665.3003669} of this paper has appeared in SIGMOD Record. In this version, we extend it to include the recent developments in this field and references to recent work.
	
	The rest of this paper is organized as follows: Section~\ref{sec:background} provides background information about deep learning models and training algorithms;  Section~\ref{sec:db2dl} discusses the application of database techniques for optimizing deep learning systems. Section~\ref{sec:dl2db} describes research problems in databases where deep learning techniques may help to improve performance. Some final thoughts are presented in Section~\ref{sec:conclusion}.

	\section{background}\label{sec:background}

	Deep learning refers to a set of machine learning models which try to learn high-level abstractions (or representations) of raw data through multiple feature transformation layers. Large training datasets and deep complex structures~\cite{DBLP:journals/corr/abs-1905-04849} enhance the ability of deep learning models for learning effective representations for tasks of interest.
	There are three popular categories of deep learning models according to the types of connections between layers ~\cite{lecun2015deep}, namely feedforward models (direct connection), energy models (undirected connection) and recurrent neural networks (recurrent connection). 
	Feedforward models, including Convolution Neural Network (CNN), propagate input features through each layer to extract high-level features. CNN is the state-of-the-art model for many computer vision tasks. Energy models, including Deep Belief Network (DBN) are typically used to pre-train other models, e.g., feedforward models.
	Recurrent Neural Network (RNN) is widely used for modeling sequential data.
	Machine translation and language modeling are popular applications of RNN.
	
	\begin{figure}
		\centering
		\includegraphics[width=.4\textwidth]{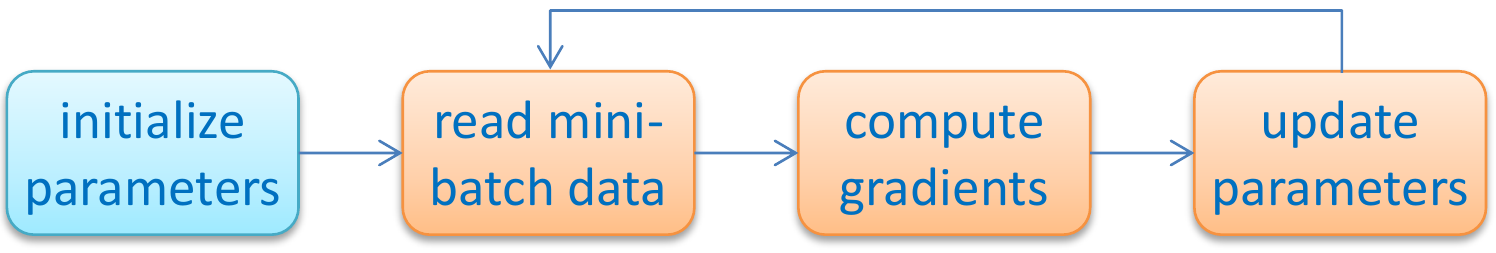}
		\caption{Stochastic Gradient Descent.}
		\label{fig:sgd}
	\end{figure}
	
\begin{figure}
		\centering
		\includegraphics[width=.3\textwidth]{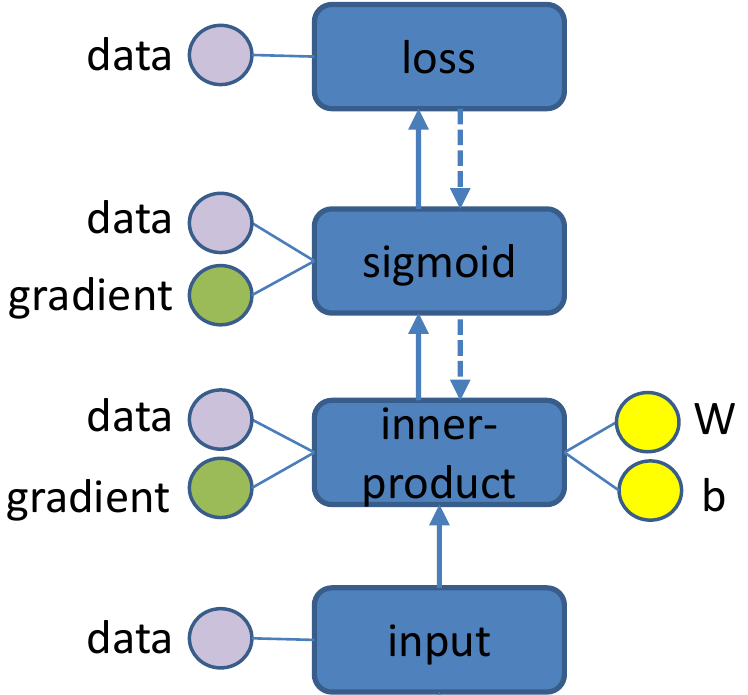}
		\caption{Data flow of Back-Propagation.}
		\label{fig:bp}
	\end{figure}

	Before deploying a deep learning model,
	the model parameters involved in the transformation layers need to be trained.
	The training turns out to be a numeric optimization procedure to
	find parameter values that minimize the discrepancy (loss function) between the
	expected output and the real output.
	Stochastic Gradient Descent (SGD) is the most widely used training algorithm.
	As shown in Figure~\ref{fig:sgd},
	SGD initializes the parameters with random values, and then iteratively refines them based on the computed gradients with respect to the loss function.
	There are three commonly used algorithms for
	gradient computation corresponding to the three model categories
	above: Back Propagation (BP), Contrastive Divergence (CD) and
	Back Propagation Through Time (BPTT).
	By regarding the layers of a neural net as nodes of a graph,
	these algorithms can be evaluated by traversing the graph
	in certain sequences.
	For instance, the BP algorithm is illustrated in Figure~\ref{fig:bp}, where a simple feedforward model is trained by traversing along the solid arrows to compute the data (feature) of each layer,
	and along the dashed arrows to compute the gradient of each layer and each parameter ($W$ and $b$).

	\section{Databases to Deep Learning}\label{sec:db2dl}
	
	In this section, we discuss the optimization techniques used in deep learning systems, and research opportunities from  the perspective of databases.

	\subsection{Stand-alone Training}
	Currently, the most effective approach for improving the training speed of deep learning models is using Nvidia GPU with the cuDNN library. Researchers are also working on other hardware, e.g. FPGA~\cite{DBLP:journals/corr/LaceyTA16}. Besides exploiting advancements in hardware technology, operation scheduling and memory management are two important components to consider.

	\subsubsection{Operation Scheduling}
\begin{figure}
		\centering
		\includegraphics[width=.45\textwidth]{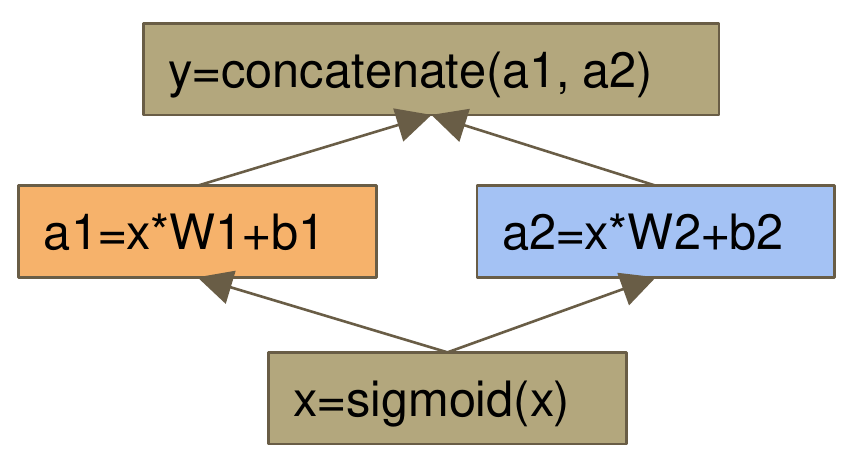}
		\caption{Sample operations from a deep learning model.}
		\label{fig:schedule}
	\end{figure}

	Training algorithms of deep learning models typically involve expensive linear algebra operations as shown in Figure~\ref{fig:schedule}, where the matrix $W1$ and $W2$ could be larger than $4096*4096$. Operation scheduling is to first detect the data dependency of operations and then place the operations without dependencies onto executors, e.g., CUDA streams and CPU threads. Taking the operations in Figure~\ref{fig:schedule} as an example, $a1$ and $a2$ in Figure~\ref{fig:schedule} could be computed in parallel because they have no dependencies. The first step could be done statically based on dataflow graph or dynamically~\cite{DBLP:journals/corr/ChenLLLWWXXZZ15} by analyzing the orders of read and write operations. Databases also have this kind of problems in optimizing transaction execution~\cite{7486988} and query plans. Those solutions should be considered for deep learning systems. For instance, databases use cost models to estimate query plans. For deep learning, we may also create a cost model to find an optimal operation placing strategy for the second step of operation scheduling given a fixed computing resources including executors and memory.
	
	\textbf{Recent developments}:  Mirhoseini et al.~\cite{46115} propose to optimize the placement of operations on heterogeneous hardware devices (e.g., CPU and GPU) using reinforcement learning. Jia et al.\cite{DBLP:journals/corr/abs-1807-05358,DBLP:journals/corr/abs-1802-04924} go beyond simple operation parallelism to consider parallelism from multiple dimensions together, including data samples and channels, operations, attributes and parameters. In addition, operation substitution has been studied in~\cite{Jia2019OPTIMIZINGDC}, which substitutes the original operations with new ones that retain the semantics but lead to better overall efficiency. Operation fusing is one example. A cost-based search algorithm is introduced to find optimized computation graphs. 
	Similar fusing techniques are applied in open-source libraries including Tensorflow~\cite{tensorflow2015-whitepaper} and PyTorch~\cite{paszke2017pytorch}.
	
	\subsubsection{Memory Management}
	Deep learning models are becoming larger and larger, and already occupy a huge amount of memory space. For example, the VGG model~\cite{DBLP:journals/corr/SimonyanZ14a} cannot be trained on normal GPU cards due to memory size constraints.
	Many approaches have been proposed towards reducing memory consumption. Shorter data representation, e.g. 16-bit float~\cite{courbariaux2014low} is now supported by CUDA. Memory sharing is an effective approach for memory saving~\cite{DBLP:journals/corr/ChenLLLWWXXZZ15}. Taking Figure~\ref{fig:schedule} as an example, the input and output of the $sigmoid$ function share the same variable and thus the same memory space. Such operations are called `in-place' operations.
	Recently, two approaches were proposed to trade-off computation time for memory. Swapping memory between GPU and CPU resolves the problem of small GPU memory and large model size by swapping variables out to CPU and then swapping back manually\cite{cui2016geeps}. Another approach drops some variables to free memory and recomputes them when necessary based on the static dataflow graph\cite{DBLP:journals/corr/ChenXZG16}.

	Memory management is a hot topic in the database community with a significant amount of research towards in-memory databases~\cite{tan2015memory,DBLP:journals/tkde/ZhangCOTZ15}, including locality, paging and cache optimization. To elaborate more, the paging strategies could be useful for deciding when and which variables to swap. In addition, failure recovery in databases is similar to the idea of dropping and recomputing approach, and hence the logging techniques in databases could be considered. If all operations (and execution time) are logged, we can then do runtime analysis without the static dataflow graph. Other techniques, including garbage collection and memory pool, would also be useful for deep learning systems, especially for GPU memory management.

	\textbf{Recent developments}: The recomputing technique has been adopted in PyTorch~\cite{pleiss2017memory}. Wang et al. \cite{wang2018superneurons} combines recomputing and swapping to optimize the memory of convolutional neural networks. Zhang et al.~\cite{DBLP:journals/corr/abs-1903-06631} propose a smart memory pool and automatic swapping strategy for deep neural networks to replace manual swapping in \cite{cui2016geeps, wang2018superneurons}. Cai et al.~\cite{DBLP:journals/corr/abs-1904-01831} propose to slice the model for reducing the memory and computational resource consumption.

	\subsection{Distributed Training}\label{sec:dist}
	Distributed training is a natural solution for accelerating the training speed of deep learning models. The parameter server architecture~\cite{Dean2012} is typically used, in which the workers compute parameter gradients and the servers update the parameter values after receiving gradients from workers. There are two basic parallelism schemes for distributed training, namely, data parallelism and model parallelism. In data parallelism, each worker is assigned a data partition and a model replica, while for model parallelism, each worker is assigned a partition of the model and the whole dataset. Two recent survey papers~\cite{Mayer2019ScalableDL,10.1145/3320060} give a comprehensive introduction about parallel and distributed deep learning. 
	The database community has a long history of working on distributed environment, ranging from parallel databases \cite{lee2000towards} and  peer-to-peer systems~\cite{vu2010peer} to cloud computing~\cite{DBLP:journals/csur/LiOOW14}. We will discuss some research problems relevant to databases arising from distributed training in the following paragraphs.

	\subsubsection{Communication}\label{sec:communication}
	Given that deep learning models have a large set of parameters, the communication overhead between workers and servers is likely to be the bottleneck of a training system, especially when the workers are running on GPUs which decrease the computation time. In addition, for large clusters, the synchronization between workers also accounts. Consequently, it is important to investigate efficient communication protocols for both single-node multiple GPU training and training over a large cluster. Possible research directions include : a) compressing the parameters and gradients before transferring~\cite{DBLP:conf/interspeech/SeideFDLY14}; b) organizing servers in an optimized topology to reduce the communication burden of each single node, e.g., tree structure~\cite{gupta2015model} and AllReduce structure ~\cite{DBLP:journals/corr/WuYSDS15} (all-to-all connection); c) using more efficient networking hardware like RDMA~\cite{DBLP:conf/icml/CoatesHWWCN13}.

	\textbf{Recent developments}: Gradient compression has shown to be effective in reducing the communication cost~\cite{Jiang:2018:SAD:3183713.3196894,Grubic2018SynchronousMD,DBLP:conf/icml/TangYLZL19,DBLP:journals/corr/abs-1712-01887,DBLP:journals/corr/WenXYWWCL17,DBLP:journals/corr/Alistarh0TV16,Jiang:2018:SAD:3183713.3196894}. Various decentralized communication frameworks~\cite{Jiang:2017:HDP:3035918.3035933,10.1145/2987550.2987586,10.5555/3327757.3327864,DBLP:journals/corr/abs-1802-05799,nccl} have been proposed to replace the centralized parameter server framework which is prone to communication bottleneck at the server side. For example, \cite{
		DBLP:journals/corr/GoyalDGNWKTJH17,
		DBLP:journals/corr/abs-1807-11205,
		DBLP:journals/corr/abs-1811-05233} use AllReduce to train large-scale networks over thousands of GPUs.

	\begin{table*}
		\centering
		\caption{Summary of optimization techniques used in existing systems as of July 18, 2016.}
		\label{tb:systems}
		\begin{threeparttable}
			\begin{tabular}{|l|l|l|l|l|l|l|}
				\hline
				\textbf{}& \textbf{SINGA} & \textbf{Caffe}\cite{jia2014caffe} & \textbf{MXNet}\cite{DBLP:journals/corr/ChenLLLWWXXZZ15} & \textbf{Tensorflow}\cite{tensorflow2015-whitepaper} & \textbf{Theano}\cite{Bastien-Theano-2012}  & \textbf{Torch}\cite{collobert2011torch7} \\ \hline
				1. operation scheduling &\checkmark&x&\checkmark&-&-&x\\\hline
				2. memory management&d+a+p&i&d+s&p&p&-\\\hline
				3. parallelism &d + m&d&d + m&d + m&-&d + m\\\hline
				4. consistency &s+a+h&s/a&s+a+h&s+a+h&-&s\\\hline
				\multicolumn{2}{l}{\footnotesize{-: unknown \quad 1. x: not available: \checkmark: available}} &
				\multicolumn{5}{l}{\footnotesize{2. d: dynamic; a: swap; p:memory pool; i: in-place operation; s: static;}}\\
				\multicolumn{2}{l}{\footnotesize{3. d: data parallelism; m: model parallelism;}}&
				\multicolumn{5}{l}{\footnotesize{4. s: synchronous; a: asynchronous; h:hybrid}}
			\end{tabular}
		\end{threeparttable}
	\end{table*}

	\subsubsection{Concurrency and Consistency}\label{sec:consistency}
	Concurrency and consistency are traditional research problems in databases. For distributed training of deep learning models, they also matter.
	Currently, both declarative programming (e.g., Theano and Tensorflow) and imperative programming (e.g., Caffe and SINGA)  have been adopted in existing systems for concurrency implementation. Most deep learning systems use threads and locks directly. Other concurrency implementation methods like actor model (good at failure recovery), co-routine and communicating sequential processes  have not been explored.
	
	Sequential consistency (from synchronous training) and eventual consistency (from asynchronous training) are typically used for distributed deep learning. Both approaches have scalability issues~\cite{singamm}. Recently, there are studies for training convex models (deep learning models are non-linear and non-convex) using a value bounded consistency model~\cite{DBLP:conf/cloud/WeiDQHCGGGX15}. Researchers are starting to investigate the influence of consistency models on distributed training~\cite{gupta2015model,DBLP:journals/corr/HadjisZMR16,DBLP:journals/corr/ChenMBJ16}. There remains much research to be done on how to provide flexible consistency models for distributed training, and how each consistency model affects the scalability of the system, including communication overhead.

	\textbf{Recent developments}: In recent papers and the benchmark testing~\cite{Coleman2017DAWNBenchA}, synchronous training is preferable to asynchronous training~\cite{DBLP:conf/icdcs/ZhangTWY18,10.5555/3060832.3060950,10.1145/2987550.2987554} because the former one is more stable in terms of convergence. With warm-up, layer-wise adaptive rate scaling for the learning rate~\cite{DBLP:journals/corr/GoyalDGNWKTJH17}, label smoothing, etc., synchronous SGD can scale to over 2000 GPUs~\cite{Yamazaki2019YetAA,DBLP:journals/corr/abs-1807-11205} without sacrificing accuracy. Typically, they increase the batch size gradually from a few thousands to tens of thousands. FlexPS~\cite{Huang2018FlexPSFP} is a system that support such  training schemes that involve multiple stages.  For very large models, pipeline training~\cite{10.1145/3341301.3359646,NIPS2019_8305} can be adopted, which partitions the model. However, data parallelism is still more popular than model parallelism since it is easier to implement and incurs less communication as well as synchronization overhead.


	\subsubsection{Fault Tolerance}\label{sec:fault}
	Databases systems have good durability via logging (e.g., command log) and checkpointing. Current deep learning systems recover the training from crashes mainly based on checkpointing files~\cite{tensorflow2015-whitepaper}. However, frequent checkpointing would incur vast overhead.
	In contrast with database systems, which enforce strict consistency in transactions, the SGD algorithm used by deep learning training systems can tolerate a certain degree of inconsistency. Therefore, logging is not a must. How to exploit the SGD properties and system architectures to implement fault tolerance efficiently is an interesting problem. Considering that distributed training would replicate the model status, it is thus possible to recover from a replica instead of checkpointing files. Robust frameworks (or concurrency model) like actor model, could be adopted to implement this kind of failure recovery.

	\subsection{Optimization Techniques in Existing Systems}
	A summary of existing systems in terms of the above mentioned optimization aspects is listed in Table~\ref{tb:systems}. Many researchers have done ad hoc optimization using Caffe, including memory swapping and communication optimization. However, the official version is not well optimized. Similarly, Torch itself provides limited support for distributed training. MXNet has optimization for both memory and operations scheduling. Theano is typically used for stand-alone training. Tensorflow  is potential for the aforementioned static optimization based on the dataflow graph.
	
	We are optimizing the Apache incubator SINGA system~\cite{opensourcesinga} starting from version 1.0. For stand-alone training, cost models are explored for runtime operation scheduling. Memory optimization including dropping, swapping and garbage collection with memory pool will be implemented. OpenCL is supported to run SINGA on a wide range of hardware including GPU, FPGA and ARM. For distributed training, SINGA (V0.3) has done much work on flexible parallelism and consistency; hence the focus would be on optimization of communication and fault-tolerance, which are missing in almost all systems.

	\section{Deep Learning to Databases}\label{sec:dl2db}\
	Deep learning applications, such as computer vision and NLP, may appear very different from database applications. However, the core idea of deep learning, known as feature (or representation) learning, is applicable to a wide range of applications. Intuitively, once we have effective representations for entities, e.g., images, words, table rows or columns, we can compute entity similarity, perform clustering, train prediction models, and retrieve data with different modalities~\cite{raey,DBLP:journals/pvldb/WangOYZZ14} etc.  We shall highlight a few deep learning models that could be adapted for database applications below.

	\subsection{Query Interface}\label{sec:interface}
	Natural language query interfaces have been attempted for decades~\cite{li2014constructing}, because of their great desirability, particularly for non-expert database users.  However, it is challenging for database systems to interpret (or understand) the semantics of natural language queries. Recently, deep learning models have achieved state-of-the-art performance for NLP tasks~\cite{DBLP:journals/corr/Goldberg15c}. Moreover, RNN has been shown to be able to learn structured output~\cite{sutskever2014sequence,vinyals2014grammar}. As one solution, we can apply RNN models for parsing  natural language queries to generate SQL queries, and refine it using existing database approaches. The challenge is that a large amount of (labeled) training samples is required to train the model. One possible solution is to train a baseline model with a small dataset, and gradually refining it with users' feedback. For instance, users could help correct the generated SQL query, and these feedback essentially serve as labeled data for subsequent training.

	\textbf{Recent developments}: Multiple annotated datasets that consist of text query and SQL query pairs have been created using templates~\cite{DBLP:journals/corr/abs-1709-00103,Cai2018AnEF} and user feedback~\cite{DBLP:journals/corr/IyerKCKZ17}. State-of-the-art solutions over the WikiSQL dataset~\cite{DBLP:journals/corr/abs-1709-00103} are listed here\footnote{\url{https://paperswithcode.com/task/text-to-sql}}. The solutions~\cite{DBLP:journals/corr/abs-1709-00103, DBLP:journals/corr/IyerKCKZ17, FineganDollak2018ImprovingTE,DBLP:journals/corr/abs-1711-04436} generally extend the sequence-to-sequence model to encode the text query and then generate the SQL query via the decoder. Domain knowledge like the SQL grammar is exploited.
	
	\subsection{Query Plans}\label{sec:plan}
	Query plan optimization is a traditional database problem. Most current database systems use complex heuristic and cost models to generate the query plan. According to \cite{haritsa2010picasso}, each query plan of a parametric SQL query template has an optimality region. As long as the parameters of the SQL query are within this region, the optimal query plan does not change. In other words, query plans are in-sensitive to small variations of the input parameters. Therefore, we can train a query planner which learns from a set of pairs of SQL queries and optimal plans to generate (similar) plans for new (similar) queries. To elaborate more, we can learn a RNN model that accepts the SQL query elements and meta-data (like buffer size and primary key) as input, and generates a tree structure~\cite{vinyals2014grammar} representing the query plan. Reinforcement learning (like AlphaGo~\cite{silver2016mastering}) could also be applied to train the model on-line using the execution time and memory footprint as the reward. Note that approaches purely based on deep learning models may not be very effective. First, the query plan is generated based on probability, which is likely to have grammar errors. Second, the training dataset may not be comprehensive to include all query patterns, e.g., some predicates could be missing in the training datasets. To solve these problems, a better approach would be combining database solutions and deep learning, e.g. using some heuristics to check and correct grammar errors.
	
	\textbf{Recent developments}: Recently, there has been an increasing trend in applying deep learning techniques for optimizing database systems, including not only query plan optimization but also data access (i.e. indexing) optimization and database configuration tuning~\cite{10.1145/3299869.3300085,10.14778/3352063.3352129}. Specifically, researchers have proposed to optimize the query plan by improving the join order selection~\cite{DBLP:journals/corr/abs-1808-03196,DBLP:conf/sigmod/MarcusP18,10.14778/3229863.3236263}, query performance prediction~\cite{10.14778/3342263.3342646}, cardinality estimation for join queries~\cite{Liu:2015:CEU:2886444.2886453,10.14778/3368289.3368296,10.1145/3299869.3320218,DBLP:journals/corr/abs-1905-06425,DBLP:journals/pvldb/YangLKWDCAHKS19}, and index recommendation~\cite{10.1145/3299869.3324957} or search~\cite{Wu2019ProgressiveNI}. Neo~\cite{10.14778/3342263.3342644} generates the query plan directly by deciding the join order, operator, and index selection together. Deep reinforcement learning is the key technique supporting these optimizations. Some challenges and possible solutions are discussed in the vision paper~\cite{DBLP:journals/corr/abs-1809-10212}. In terms of data access optimization, Kraska et al.\cite{Kraska:2018:CLI:3183713.3196909} propose a learned index for read-only, in-memory database systems. It uses neural networks to map the key to the location of the record. Subsequent works have extended it for multi-dimensional index~\cite{nathan2019learning}, updatable index~\cite{ding2019alex}, dynamic workloads~\cite{DBLP:journals/corr/abs-1902-00655}, and accessing data on disk~\cite{DBLP:journals/pvldb/KakaraparthyPPK19,10.1145/3299869.3319860,ferragina2019pgmindex}. To go one step further, SageDB~\cite{Kraska2019SageDBAL} puts forth a vision where every component (such as query plan, data access, and query execution~\cite{DBLP:journals/corr/abs-1907-08817}) of a database system is optimized via machine learning models against the data distribution and (query) workload~\cite{10.1145/3299869.3314034}.

	\subsection{Crowdsourcing and Knowledge Bases}
	Many crowdsourcing~\cite{DBLP:journals/pvldb/Wu0HZL15} and knowledge base~\cite{DBLP:journals/pvldb/DongGHHMSZ14} applications involve entity extraction, disambiguation and fusion problems, where the entity could be a row of a database, a node in a graph, etc. With the advancements of deep learning models in NLP~\cite{DBLP:journals/corr/Goldberg15c}, it is opportune to consider deep learning for these problems. For example, we can learn representations for entities and then do entity relationship reasoning ~\cite{socher2013reasoning} and similarity calculation.
	
	\textbf{Recent developments}: DeepER~\cite{DBLP:journals/corr/abs-1710-00597} exploits LSTM models to learn tuple embedding for entity resolution. IDEL~\cite{8679486} implements neural entity-linking in MonetDB, where the entities are embedded. Deep learning models like CNN and attention modelling have been applied for concept linking~\cite{francis-landau-etal-2016-capturing, Dai:2018:FCL:3183713.3196907}. Mudgal et al.\cite{Mudgal:2018:DLE:3183713.3196926} evaluate four different deep learning models for entity matching problems. This website\footnote{\url{https://github.com/thunlp/KRLPapers}} keeps track of the recent papers~\cite{8047276} on knowledge representation learning and embedding.

	\subsection{Spatial and Temporal Data}
	
	Spatial and temporal data are common data types in database systems~\cite{guo2014towards}, and are commonly used for trend analysis, progression modeling and predictive analytics. Spatial data is typically processed by mapping moving objects into rectangular blocks. If we regard each block as a pixel of one image, then deep learning models, e.g., CNN, could be exploited to extract the spatial locality between nearby blocks. For instance, if we have the real-time location data (e.g., GPS data) of moving objects, we could learn a CNN model to capture the density relationships of nearby areas for predicting the traffic congestion for a future time point. When temporal data is modeled as features over a time matrix, deep learning models, e.g. RNN, can be designed to model time dependency and predict the  occurrence in a future time point.  A particular example would be disease progression modeling~\cite{CPT:CPTCLPT201253} based on historical medical records, where doctors would want to estimate the onset of certain severity of a known disease. In fact, most healthcare data is  time-serise data, and thus deep learning can make great contribution in healthcare data analysis~\cite{Lee2017,DBLP:conf/icde/LuoCGZN0L18}.
	
	\textbf{Recent developments}: Deep learning models including CNN and RNN have been applied in various spatial-temporal problems, including traffic flow prediction~\cite{Ma2019TrafficPredictTP,Jiang2018DeepUrbanMomentumAO}, travel time estimation~\cite{Wang2018WhenWY,Li2018MultitaskRL,Wang2018LearningTE}, driver behavior analysis~\cite{Wang2018YouAH}, geospatial aggregation querying~\cite{2019arXiv190606085V}, etc. A comprehensive survey of the  recent progress of applying deep learning for spatial-temporal data is presented in \cite{DBLP:journals/corr/abs-1906-04928}.

	\section{Conclusions}\label{sec:conclusion}
	In this paper, we have discussed databases and deep learning. Databases have many techniques for optimizing system performance, while deep learning is good at learning effective representation for data-driven applications. We note that these two ``different" areas share some common techniques for improving the system performance, such as memory optimization and parallelism. We have discussed  some possible improvements for deep learning systems using database techniques, and research problems applying deep learning techniques in database applications. 
	To make the database systems more autonomic, with the ability to learn and optimize, and to support complex analytics and predictions beyond data aggregation, we foresee a seamless integration of ML/DL and database technologies. With the implementation of 5G mobility network, we foresee the distribution of databases, training and inference at the edge devices, which will lead to further integration and adaptation of technologies.
	Let us not miss the opportunity to contribute to the existing challenges ahead!
	
	\section{Acknowledgement}
	We would like to thank Divesh Srivastava for his valuable comments. This work was supported by  the  National  Research
	Foundation,  Prime  Minister's  Office,  Singapore,  under  its
	Competitive Research Programme (CRP Award No. NRF-CRP8-2011-08), and Singapore Ministry of Education Academic Research Fund Tier 3 under MOE’s official grant number MOE2017-T3-1-007.
	Meihui Zhang was supported by China Thousand Talents Program
	for Young Professionals (3070011181811). 
	
	\bibliographystyle{abbrv}
	{\small
		\bibliography{paper}
	}
\end{document}